# Foundations and Design Principles of Lightweight Cryptography for IoT Systems

Arsalan Vahi[1][0000-0001-7412-8523]
[1]Sabanci University, Istanbul, Turkey
arsalan.vahi@sabanci.univ.edu

**Abstract.** The successful deployment of the Internet of Things (IoT) applications relies heavily on their robust security, and lightweight cryptography is considered an emerging solution in this context. While existing surveys have been examining lightweight cryptographic techniques from the perspective of hardware and software implementations or performance evaluation, there is a significant gap in addressing different security aspects, such as design principles, specific to the IoT environment. This study aims to bridge this gap. This research presents an examination with focusing on the security evaluation of symmetric lightweight ciphers commonly used in IoT systems. The objective of this study is to provide a concise overview of lightweight ciphers with emphasizing on their security challenges which is an essential consideration for real-time and resource-constrained applications.

## 1 Introduction

The rapid development of IoT has revolutionized the global digital ecosystem. It enables integrated communication between devices. Its applications have proliferated across domains, including smart homes, healthcare monitoring, industrial automation, and smart transportation. Due to the sensitivity of their data and pervasive usage, ensuring their security is very important. This is particularly different, especially for resource-limited environments where traditional encryption algorithms cannot be applied due to computational limitations.

While lightweight cryptography emerged as a promising way for securing IoT devices, most current studies evaluate ciphers based on functional metrics such as execution speed and energy consumption rather than conducting a comprehensive assessment of their security properties. This study examines the foundation and design principles of lightweight ciphers. Several surveys have been published with the aim of examining lightweight ciphers. Some studies examined software implementations [1,2], while hardware design has been examined in [3,4]. Throughput issues have been addressed in [5], and the security aspects of lightweight ciphers are examined in [6,7]. Existing surveys on lightweight ciphers have largely focused on functional criteria such as execution time and energy consumption, while often neglecting thorough security

evaluations—particularly with respect to foundation and design principles of lightweight ciphers that can be utilized in IoT devices.

This paper addresses this research gap by conducting an in-depth security assessment of lightweight ciphers, with a special emphasis on design principles as a critical parameter. The main contributions of this study include providing a precise definition of the Internet of Things and its key components, classifying low-resource (or resource-constrained) devices and outlining IoT security challenges and emphasizing the role of lightweight cryptography in addressing them.

The structure of this article is as follows: Sections 2 and 3 reviews the background on IoT and resource-constrained devices. In Section 4, the fundamentals of lightweight cryptography, including its design strategies and distinguishing features, the security challenges and concerns associated with IoT are explored.

# 2 Background

## 2.1 Internet of Things

The concept of the Internet of Things (IoT) was first introduced by Kevin Ashton in 1999 [8]. IoT has emerged from the convergence of five key mature technologies: ubiquitous computing, sensor networks, big data, cloud computing, and radio frequency identification (RFID) technology. Fundamentally, IoT refers to a networked ecosystem of objects embedded within an environment, interconnected through a dynamic public network [8, 9]. These objects are capable of self-configuration and communication using standard, universally accepted protocols. They gather environmental data, transmit it over the Internet to centralized or decentralized data analysis centers, and receive intelligent, context-aware services in return.

Based on the function of key components within the IoT ecosystem, they can be categorized into two broad groups: platforms/services and hardware devices. Figure 1 shows this categorization with their components.

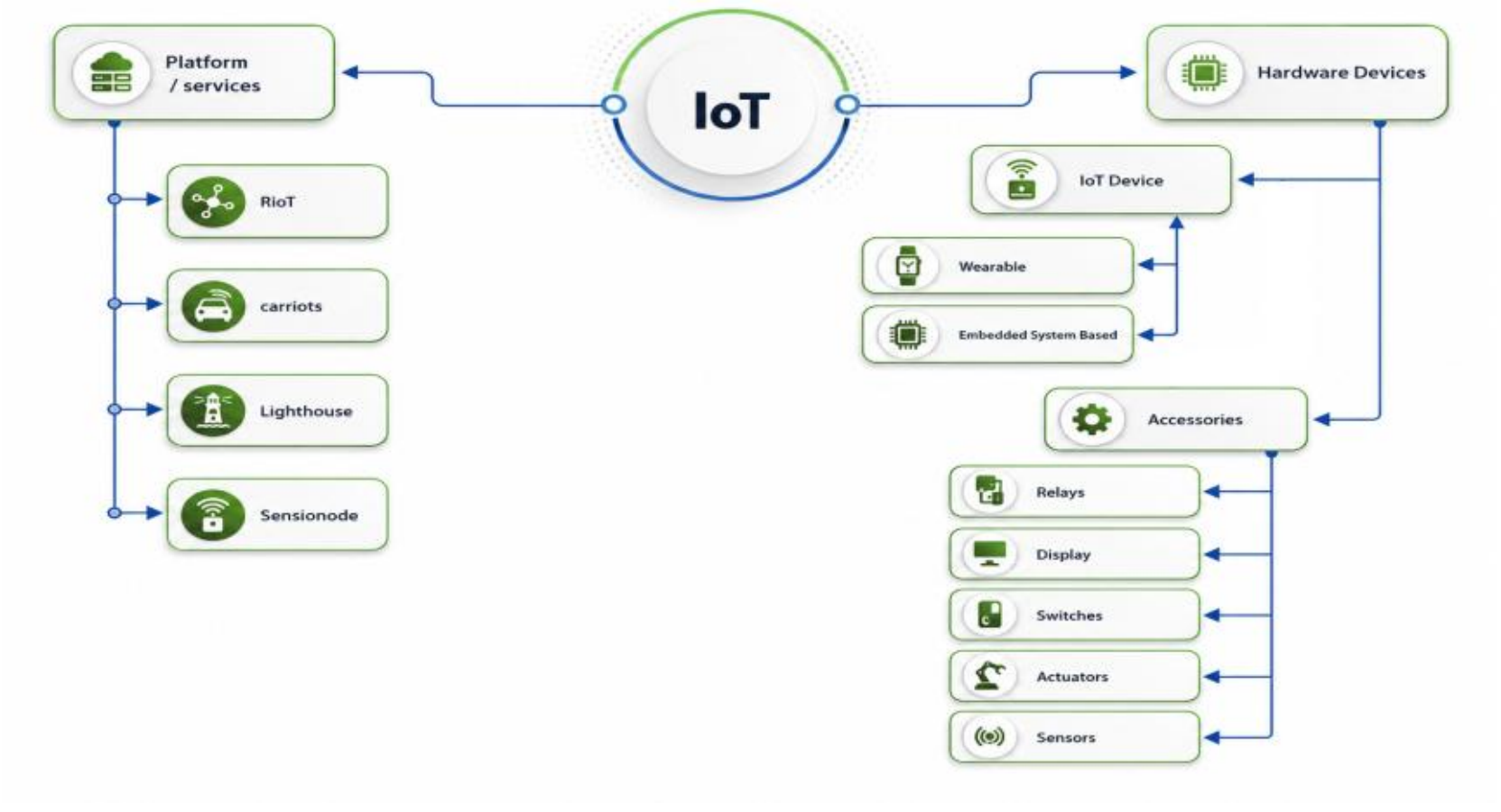

**Fig 1.** Key components of IoT

## 3 Low-Resource Devices

In another classification, IoT devices can generally be categorized into two broad groups, CPU-based devices and microcontroller-based devices.

•**CPU-based devices** are often referred to as non-constrained or low-constrained devices. These devices face few, if any, significant resource limitations. They typically employ powerful CPUs, high memory capacity, stable power supplies, and high-speed, reliable connectivity. Smartphones, laptops, edge servers, Raspberry Pi (in certain configurations), and cloud-connected gateways are examples of CPU-based devices. Such devices usually serve as hubs, data aggregators, or gateways within IoT networks and can perform complex tasks such as data encryption, analytics, and decision-making.

•**Microcontroller-based devices** are also known as resource-constrained or low-resourced devices. These devices have limited computational capacity. They operate on low CPU power, limited RAM and ROM, low bandwidth, and are often battery powered. Sensor nodes, RFID tags, wearable devices, and microcontrollers such as the Atmega328p (Arduino) and ESP8266 are well-known examples of these devices. These devices are widely used in IoT environments due to their small size, low cost, and energy efficiency [9].

These two general categories can be further classified into three distinct device classes based on their hardware capabilities [10]:

**Class 0 – Low-end IoT Devices**: These devices have extremely limited resources in terms of memory, power, and processing capacity. Their main function is sensing and transmitting data using lightweight communication protocols. In some research, ultra lightweight ciphers are presented as a solution.

**Class 1 – Mid-range IoT Devices**: Class 1 devices employ microcontrollers with more computational resources than Class 0. Their role is serving as enhanced nodes in IoT networks.

**Class 2 – High-end IoT Devices (Single Board Computers)**: These devices are equipped with substantial processing power, memory, and storage. These resources make suitable for running full-fledged operating systems like Linux and Windows. In Table 1 summary of these classes as well as some other characteristics is described [9].

Table 1. classification of low resource devices

| Class # | Name | Clock | RAM | Flash Memory | Processing Unit | Security concerns | built-in security | Encryption | Applications |
|---|---|---|---|---|---|---|---|---|---|
| Class 0 | Low-end IoT devices | Less than 100 MHz | 1-50 KB | 10-50 KB | 8/16/32-bit architecture | High | No | Lightweight/ ultra lightweight | Sensing/ Actuating e.g., Sensors |
| Class 1 | Middle-End IoT devices | 100 Mhz-1.5 GHz | 100 KB-100 MB | 10KB-100MB | 8/16/32-bit architecture | High | No | Lightweight/ conventional | Image recognition e.g., Arduino |
| Class 2 | Mostly Single-Board computers | --- | --- | --- | 8/16/32/64-bit architecture | High | Yes/No | Lightweight/ conventional | Machine Learning e.g., Raspberry pi |

Security concerns for low-resourced devices are closely tied to the sensitivity of the data they handle and the criticality of the applications where they are deployed. As an example, when comparing the sensors in Wireless Sensor Network (WSN) to Wireless

Body Area Networks (WBAN), the challenges faced with the latter one could be more complicated. Regarding WBAN sensors, the primary security requirements should include ensuring the confidentiality, integrity, and accuracy of the transmitted data.

## 4 Lightweight Cryptography

Cryptography, which is often known as secret writing, plays a paramount role in securing data and communications in a vast spectrum of hardware and software platforms. This section briefly addresses the applications and importance of lightweight cryptography. In addition, it provides an overview of the key applications and significance of lightweight cryptography

### 4.1 Role of Lightweight Cryptography

Most organizations rely on cryptographic techniques to enhance the security of their systems. However, the application of conventional cryptographic algorithms is often unsuitable for systems with resource constraints. While Class 1 and Class 2 devices typically have sufficient computational power, memory, and energy to support standard cryptographic methods, many heterogeneous IoT devices operate under severe limitations—such as restricted processing capability, low memory, limited power supply, short battery life, and constrained storage capacity. In such cases, lightweight cryptography emerges as a necessary alternative. These lightweight solutions are expected to deliver security services equivalent to conventional cryptographic algorithms while remaining feasible for implementation in constrained environments.
Importantly, applying conventional cryptography in low-resource settings can introduce significant challenges, including performance degradation and excessive power consumption. The fundamental principle guiding the design of lightweight ciphers is reduced complexity. This makes them better suited to limited-resource devices. Unlike conventional ciphers, which are not optimized for such environments, lightweight cipher design emphasizes a trade-off between robust security and efficient functionality. Key considerations include minimizing power and energy consumption, optimizing battery usage, reducing memory requirements, lowering performance overhead, and maintaining strong security guarantees. Among these, low energy and power consumption are essential criteria for categorizing a cipher as lightweight.

### 4.2 Lightweight cryptographic solutions for IoT devices

Existing research identifies three primary approaches to designing lightweight ciphers [11].

The first approach involves compact hardware implementations of standard and trusted conventional block ciphers. The focus here is on achieving optimized, resource-efficient versions of existing algorithms. The second approach involves slightly modifying well-known and widely used standard ciphers. The third approach is the design of entirely new low-cost block ciphers and stream ciphers, specifically tailored for constrained environments. This strategy aims to achieve a careful balance between security, cost, and performance.

Table 2: Design approach for different lightweight ciphers [11]

| Approach | Ciphers | Features |
|---|---|---|
| Compact Hardware Implementations for Standard Block Ciphers | Low-cost AES (128,192,256-bit key) | - 8-bit Datapath is used for the round operation<br>- On-the-fly key expansion<br>- S-Box is implemented as combinatorial logic<br>- 3,400 to 3,100 GE requirements<br>- Able to encrypt 128-bit data within 1,032 and 160 clock cycles |
| | XTEA (128-bit key) | - Can encrypt 64-bit block within 112 clock cycles<br>- Requires 3,490 GE (ASIC) and 254 slices (FPGA) |
| Slight Modification of a classical Block cipher | DESL (184-bit key) | - Replace the eight s-boxes with a single new one.<br>- 1,848 GE requirements<br>- Encrypts on 64-bit data block within 144 clock cycles |
| New low-cost block ciphers and stream ciphers | Hight (128-bit key) | - 3,048 GE |
| | mCrypton (64,96, 128-bit) | - 2,420 GE requirements |
| | Present (80, 128-bit key) | - 1,000 GE requirements |

Table 2 presents representative examples for each of these approaches, along with their key characteristics and known attacks. In Table 3 key design considerations in lightweight cipher design are briefly provided.

**Table 3.** Design key considerations for a lightweight cipher

| Consideration | Goals |
|---|---|
| Functionality | - The goal of the designed service is to provide multifaceted services within a single cipher |
| Design Objectives | - Designers must meet some special objectives. For instance, some ciphers - may be tailored for devices with limited message lengths, necessitating a focus on efficiency in handling shorter data. |
| Design Category | - Designers mostly intended to design ciphers that fall within one of the designated categories. |
| Performance Characteristics | - Factors related to performance, including latency, throughput, and power consumption, should be thoughtfully considered. |
| Security Characteristics | - Security considerations encompass a range of factors, including key size, number of rounds, relevant attack models, cryptographic strength, and resistance to implementation attacks such as side-channel attacks. Cipher designers must strike a delicate balance between these security aspects. |

Meeting all key design considerations in the development of a cipher presents a significant challenge. In most cases, designers must prioritize these considerations, often make trade-offs or adapt them to suit specific application requirements. However, this prioritization can sometimes lead to the neglect of critical security properties—even though ensuring robust security is the fundamental objective in the design of any lightweight cipher.

## 4.3 Functionality considerations for designing a lightweight cipher

Lightweight ciphers are engineered to operate efficiently across a broad spectrum of applications, ranging from low-resource devices such as sensors and RFID tags to more capable platforms like smartphones, laptops, and desktop computers.

The common classification of lightweight ciphers based on functionality aligns closely with that of conventional ciphers. This classification considers the type of cryptographic operations the ciphers perform and their intended use cases. In this context, the primary focus is on the purpose or role of the cipher, rather than the hardware platform it is designed for.

According to functionality, lightweight ciphers are typically categorized into six distinct classes of Block ciphers, Stream ciphers, Authenticated Encryption (AE), Hash Functions, Message Authentication Codes (MACs) and Public-key lightweight cryptography.

## 4.4 Design objective considerations for designing a lightweight cipher

Based on their design objectives, lightweight encryption can be broadly categorized into two domains of **Low-End Hardware** which encompasses devices like RFID tags and smart cards, which typically have very limited computational capabilities and **Real-Time Software Devices** which includes devices equipped with 8, 16, or 32-bit embedded microprocessors and focuses on real-time software-based cryptography.

It is important to note that lightweight ciphers typically fall into one of the previously mentioned categories of design objectives, and no single cipher can simultaneously satisfy the requirements of both categories. As a result, selecting a suitable cipher depends on its intended application domain, and there is no lightweight cipher that offers the same level of versatility as conventional ciphers.

One critical aspect that cipher designers must consider is the architectural diversity of microcontrollers, which can vary in word size (e.g., 4-bit, 8-bit, 16-bit, and 32-bit), each with its own specific Instruction Set Architecture (ISA). This architectural variation implies that not all operations can be executed in a single clock cycle, and only a limited set of basic instructions may be supported. For instance, if the ISA supports bitwise operations such as XOR, AND, and OR, but a bit-substitution operation is required, it may not be achievable in a one-clock cycle. This limitation can lead to longer execution times and increased energy consumption.

In the case of RFID devices, many of which operate without internal batteries and instead harvest energy from their environment—energy efficiency becomes an even more critical concern. Therefore, the selection of a lightweight cipher must not only be based on device constraints but should also account for environmental factors and the nature of the application.

In other words, a well-designed lightweight cipher should be optimized for constrained environments, such as Class 1 devices like sensors and smartcards, while also being adaptable to non-constrained platforms. Ultimately, the choice between lightweight and conventional cryptographic algorithms should be informed by the specific limitations of the application, the characteristics of the target device, and the operational requirements of the deployment environment.

### 4.5 Performance characteristics considerations for designing a lightweight cipher

In designing a lightweight cipher, especially those aimed at performing in limited-resourced devices like RFID tags considering performance characteristics are critical. The designer must balance the security, efficiency, and required resources. In Table 4, the key performance characteristics are listed.

**Table 4**: Performance concerns and their related characteristics and Goals

| Performance concerns | Characteristics and Goals |
|---|---|
| Hardware efficiency | Gate Equivalents (GE): A measurement of hardware footprint (typically under 2000 GEs is desirable) / Latency: Time delay (measured in clock cycles) between the start of encryption and the output of the first result. Low latency is desired. |
| Software Efficiency | Code Size: A smaller binary is desirable for limited memory / RAM/ROM usage: Minimum usage of both RAM/ROM / Cycle Count: Total number of cycles needed for one full encryption or decryption; Lower cycle counts mean faster execution and less power usage / Throughput: Amount of data processed per unit time (e.g. bit/sec) / Latency: Time in cycles until the first output is ready. It is a matter in time-sensitive systems such as automative or industrial controls. |
| Energy and Power Consumption | Consumption of minimal energy per bit processed / Energy-efficient design avoids complex operations like large S-boxes and multiplication in finite fields. |
| Speed | Important for real-time applications or systems that handle large amounts of data |
| Flexibility | Compatible with different hardware and software platforms. |
| Implementation | Easy to implement and verify. |
| Simplicity | Simple round functions and key schedules are easier to implement and verify. |
| Support for Parallelism | May be useful for enhancing throughput in certain applications |

## 5 Security Challenges and Considerations for Designing a Lightweight Cipher for IoT

On the Internet of Things (IoT), device connectivity is primarily established through networks and the Internet. One of the fundamental challenges in such interconnected systems is the risk of unauthorized access. IoT devices—often referred to as "things"—collect, process, and transmit valuable information, making security a critical concern. In the context of IoT, security refers to safeguarding devices and their associated networks against threats and attacks commonly encountered on the Internet. This involves the identification, monitoring, and mitigation of potential security breaches. Given that IoT data can be sensitive and subject to theft or leakage, any compromise can have severe consequences, especially since IoT systems influence both the physical and virtual dimensions of their environments. Consequently, ensuring the security of IoT systems is a high priority for both end-users and IT professionals.

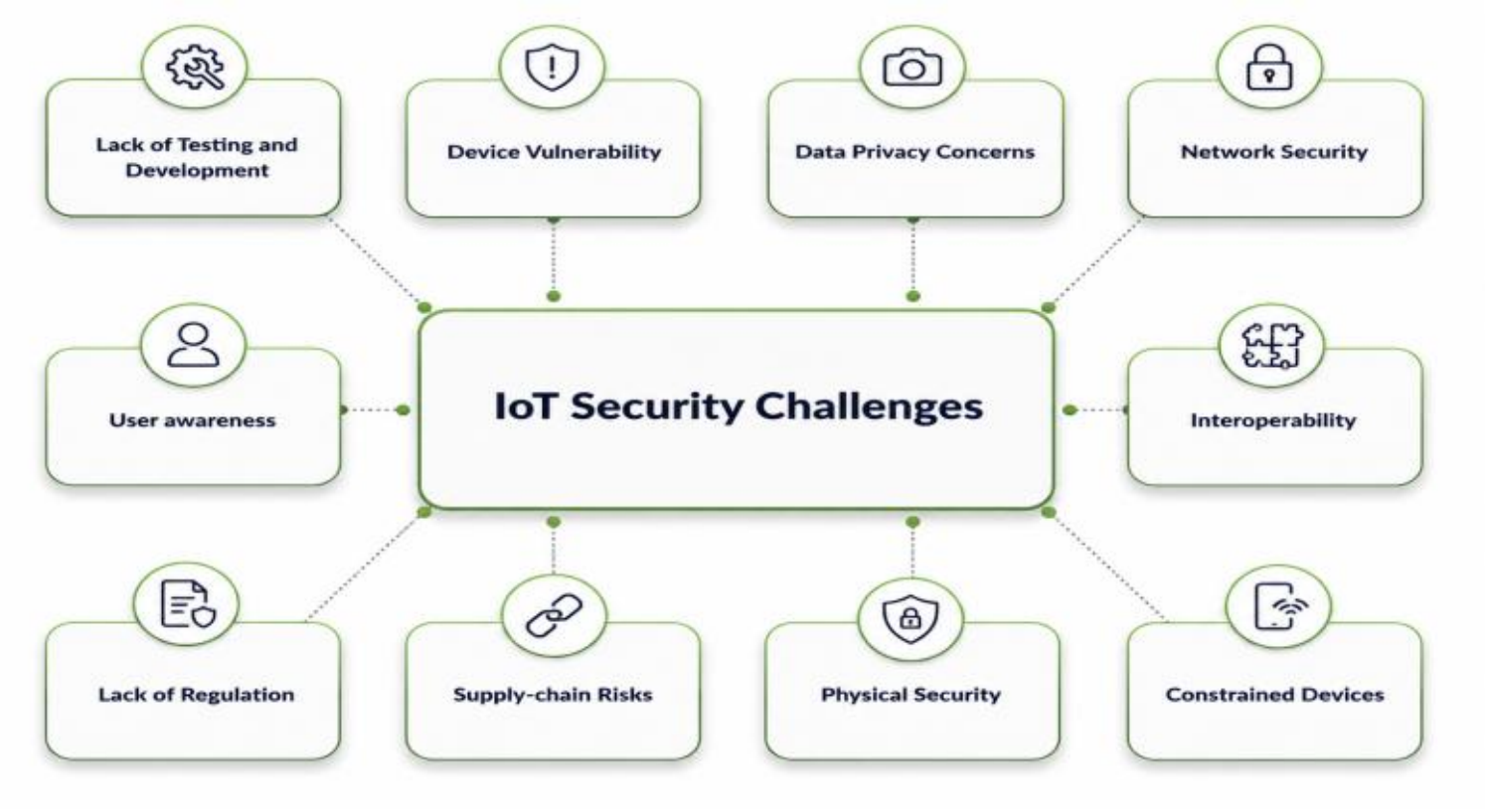

**Fig 2.** IoT security challenges

As the number of IoT-connected devices continues to grow, network vulnerabilities present increasingly attractive targets for cybercriminals. This expansion brings forth a range of security challenges unique to the IoT domain. Fig. 2 illustrates some of the key security challenges faced in IoT environments.

Some IoT manufacturers rush to bring products to market and treat security as an afterthought; as a result, device-level security is often neglected, and once deployed, these devices may lack essential protection. Given that most IoT devices possess limited computational capabilities, they are particularly vulnerable to attacks. Weak or default passwords, unpatched firmware, and the absence of robust security mechanisms expose such devices to exploitation [12].

**Table 5.** Security considerations related to resource constraint devices

| Security Concerns | Goals |
|---|---|
| Limited Computing Power | These devices are equipped with microcontrollers or low-power processors. This impacts the ability to implement complex security algorithms. |
| Limited Memory Usage | These devices often have low memory for storing security-related data such as encryption keys and security certificates. Security measures must be memory-intensive. |
| Energy Efficiency | These devices are commonly battery-powered or rely on energy harvesting mechanisms (such as RFID). Energy efficiency is very important and there should be a reasonable balance between security and energy consumption. |
| Real-time Constraints | Some IoT applications require real-time processing and decision making such as autonomous vehicles. Security mechanisms must be designed to operate within real-time constraints. |
| Cost Considerations | As IoT devices are often deployed in large numbers, cost dictates that this product should be affordable. Even, in some cases, the products are manufactured with no security to decrease the cost. |
| Scalability Challenges | Managing security across numerous devices is complex. Resource constraints can exacerbate these scalability challenges. |
| Security vs. Functionality | There should be a balance between security requirements and the primary functionality of IoT devices. Resource limitations may force trade-offs between security and desired features. |

Moreover, IoT devices collect and transmit large volumes of data, raising significant privacy concerns. Unauthorized access and data breaches can have serious implications for both individuals and organizations [13]. The ecosystem of IoT often comprises heterogeneous devices from various manufacturers, leading to interoperability issues that may compromise security.

Physical access to IoT devices increases the risk of tampering and theft of sensitive information. Additionally, vulnerabilities introduced through compromised software or components during the manufacturing process constitute a major supply chain risk. Keeping up with evolving security regulations and standards remains an ongoing challenge for developers and manufacturers [14].

Educating end-users on best practices for IoT security is crucial, as user ignorance can inadvertently contribute to security vulnerabilities. Furthermore, constrained devices may be incapable of supporting conventional security protocols due to their limited processing power, memory, or energy resources. This presents a unique challenge for both manufacturers and users. Table 5 outlines key security considerations for resource-constrained IoT devices.

Also, regarding some specific low-constrained devices such as WBAN, more challenges are considered including [15]:

1. **Power/Energy Consumption:** WBAN sensors are often constrained by strict energy consumption limitations due to their deployment on or within the human body. This energy constraint makes it imperative to employ security mechanisms that minimize power consumption, which may differ from the strategies employed in WSNs.
2. **Battery Replacement:** In WBANs, particularly in medical and healthcare applications, frequent battery replacement or recharging may not be feasible. Therefore, security solutions should be designed to maximize the lifespan of the device's battery.
3. **Thermal Radiation:** WBAN devices near the human body generate thermal radiation, which can affect the device's operation and longevity. Security mechanisms must account for these thermal factors.
4. **Number of Sensors:** The number of sensors utilized in WBANs may vary significantly depending on the specific medical or healthcare application. WBANs have a smaller number of sensors compared with generic WSNs. This dynamic nature necessitates adaptable and scalable security solutions.

# 6 Conclusion

As a result, in order to target security methods for low-resourced devices, it should be considered that directly adopting existing methods is not suitable and some factors should be considered and should not be directly borrowed from existing security approaches. Instead, it is crucial to develop and adopt lightweight security schemes that are tailored to the specific requirements and constraints of different applications.